# Principal component analysis within nuclear structure

A. Al-Sayed[*]

*Physics Department, Faculty of Science, Zagazig University, Zagazig, Egypt.*

*Physics Department, College of Sciences and Arts in Muthnib, Qassim University, KSA.*

## Abstract

The principal component analysis (PCA) of different parameters affecting collectivity of nuclei predicted to be candidate of the interacting boson model dynamical symmetries are performed. The results show that, the use of PCA within nuclear structure can give us a simple way to identify collectivity together with the parameters simultaneously affecting it.

## 1. Introduction

There are two main problems facing nuclear physicists; the lack of knowledge of the exact nature of nuclear force, and the variety of nuclear structures. Many trials are developed to classify the collective behavior of atomic nuclei specially the even-even ones. The interacting boson model (IBM)[1] is a good candidate on this way.

The IBM introduced what is known as dynamical symmetries, which are usually referred to the *U(5)*, *O(6)* and *SU(3)* limits. These limits are corresponding to the familiar spherical, *γ*-unstable, and well-deformed nuclei respectively. Several trials have been considered to introduce an order parameter to identify nuclei belonging to each dynamical symmetry. Among these is the ratio $R_{4/2}$ [2,3] of excitation energies of the first $4^+$ and the first $2^+$ excited states. The IBM calculation of energy levels yields values of $R_{4/2}$ = 2.00, 2.50, and 3.33 for the dynamical symmetries *U(5)*, *O(6)*, and *SU(3)*, respectively.

Beside the $R_{4/2}$ ratio, numerous different variables are proposed to facilitate the identification of collective behavior through atomic nuclei. Among these variables; number of protons (Z), quadrupole deformation parameter ($β_2$) [4], P-factor [5-7], and reduced electric quadrupole transition rate from $0^+$ ground state to the first excited $2^+$ state B(E2; $0_1^+ \rightarrow 2_1^+$) or simply B(E2)↑ [8].

Most studies focused on classifying nuclei according to a certain variable to facilitate the conclusions. The question is now, what is the effect if we take into consideration all these variables on nuclear structure simultaneously? In other words, do even-even nuclei share the same degree of correlation with these variables? To answer this question, we turn to multivariate analysis in standard statistics.

We describe the data set in section 2. A brief account of the method of analysis is given in section 3. In section, 4 results and discussions are presented. The conclusion of this work is outlined in Section 5.

*- email address: a.alsayed@zu.edu.eg

## 2- Data set

Even-even nuclei are chosen for their abundance and simplicity to be described by nuclear models. The data set is taken from National Nuclear Data Center [9] to calculate the $R_{4/2}$ ratio, while $\beta_2$ data from [10], and B(E2)↑ values from a recent update [11] in $e^2b^2$ units. Nuclei having adopted B(E2)↑ are only taken into consideration. We restrict our study to nuclei predicted by the $R_{4/2}$ ratio to represent dynamical symmetries of IBM. We put this condition to simplify the biplot chart as given in the results.

Practically, the $R_{4/2}$ is not a well "order parameter" to distinguish varieties within nuclear structure. The $R_{4/2}$=2.50 does not correspond only to γ-unstable nuclei, instead, it corresponds to different structures spanning between O(6) to critical point of a spherical to axially deformed transition, U(5)-SU(3). Consequently, we will exclude that value 2.5 of the ratio $R_{4/2}$ from our calculations. The range of 2.05≤ $R_{4/2}$ < 2.30 will be taken into consideration through current study since most empirically anharmonic vibrational nuclei found within that ratio. [12, 13]. In this way, we get 10 nuclei in the range 1.95≤ $R_{4/2}$ < 2.05, 47 nuclei in the range 2.05≤ $R_{4/2}$ < 2.30, and 41 nuclei of $R_{4/2}$≥ 3.25. The data set is shown in table (1).

| Nucleus | B(E2) | $P$ | $Z$ | $\beta_2$ | $2^+$ | $R_{4/2}$ | Nucleus | B(E2) | $P$ | $Z$ | $\beta_2$ | $2^+$ | $R_{4/2}$ |
|---|---|---|---|---|---|---|---|---|---|---|---|---|---|
| **Sr90**  | 0.102  | 1.6667 | 38 | 0.053  | 0.8317 | 1.991  | **Se68**  | 0.216  | 3      | 34 | 0.24   | 0.8542 | 2.2735 |
| **Te118** | 0.57   | 1.7778 | 52 | -0.147 | 0.6057 | 1.9917 | **Ge64**  | 0.208  | 2      | 32 | 0.219  | 0.9017 | 2.2764 |
| **Ni62**  | 0.0887 | 0      | 28 | -0.096 | 1.1729 | 1.9919 | **Ba136** | 0.413  | 1.5    | 56 | 0      | 0.8185 | 2.2805 |
| **Hg186** | 1.37   | 1.8182 | 80 | -0.13  | 0.4053 | 1.9934 | **Zn62**  | 0.1224 | 1.3333 | 30 | 0.209  | 0.954  | 2.2915 |
| **Rn218** | 0.89   | 2.4    | 86 | 0.04   | 0.3242 | 2.0144 | **Cd112** | 0.481  | 1.75   | 48 | 0.144  | 0.6175 | 2.2924 |
| **Te128** | 0.38   | 1.5    | 52 | 0      | 0.7432 | 2.0147 | **Pd102** | 0.46   | 2.4    | 46 | 0.143  | 0.5564 | 2.2931 |
| **Zn70**  | 0.152  | 1.6667 | 30 | 0.045  | 0.8844 | 2.0198 | **Nd146** | 0.748  | 2.8571 | 60 | 0.161  | 0.4538 | 2.2966 |
| **Zr88**  | 0.086  | 1.6667 | 40 | 0.053  | 1.057  | 2.0236 | **Cd114** | 0.5357 | 1.7778 | 48 | 0.163  | 0.5585 | 2.2987 |
| **Te126** | 0.4738 | 1.6    | 52 | -0.105 | 0.6663 | 2.0432 | **Sm154** | 4.355  | 5.4545 | 62 | 0.27   | 0.082  | 3.2546 |
| **Xe134** | 0.336  | 1.3333 | 54 | 0      | 0.847  | 2.0439 | **W180**  | 4.419  | 5.7143 | 74 | 0.258  | 0.1035 | 3.26   |
| **Dy152** | 0.43   | 3.2    | 66 | 0.153  | 0.6138 | 2.0547 | **Nd152** | 4.1    | 5      | 60 | 0.262  | 0.0725 | 3.2634 |
| **Te108** | 0.387  | 1.5    | 52 | 0.134  | 0.6252 | 2.0617 | **Yb168** | 5.57   | 6.8571 | 70 | 0.284  | 0.0877 | 3.2663 |
| **Sr86**  | 0.1341 | 1.6667 | 38 | 0.053  | 1.0767 | 2.0709 | **Hf174** | 5.12   | 6.6667 | 72 | 0.285  | 0.091  | 3.2685 |
| **Te124** | 0.58   | 1.6667 | 52 | -0.113 | 0.6027 | 2.0717 | **Dy160** | 5.08   | 6.8571 | 66 | 0.272  | 0.0868 | 3.2703 |
| **Ge70**  | 0.1681 | 2.8571 | 32 | -0.241 | 1.0393 | 2.0722 | **Th230** | 8.12   | 5.0909 | 90 | 0.198  | 0.0532 | 3.2726 |
| **Ge72**  | 0.2068 | 2.8571 | 32 | -0.224 | 0.834  | 2.0723 | **W184**  | 3.702  | 5.3333 | 74 | 0.24   | 0.1112 | 3.2736 |
| **Te120** | 0.709  | 1.75   | 52 | -0.156 | 0.5604 | 2.0728 | **U230**  | 9.5    | 5.4545 | 92 | 0.199  | 0.0517 | 3.2742 |
| **Mo96**  | 0.2775 | 2.6667 | 42 | 0.08   | 0.7782 | 2.0922 | **Er164** | 5.5    | 7      | 68 | 0.273  | 0.0914 | 3.2768 |
| **Te114** | 0.556  | 1.7143 | 52 | 0.161  | 0.7089 | 2.0937 | **Th232** | 9.11   | 5.3333 | 90 | 0.207  | 0.0494 | 3.2838 |
| **Te122** | 0.65   | 1.7143 | 52 | -0.139 | 0.5641 | 2.0943 | **Hf176** | 5.48   | 6.875  | 72 | 0.277  | 0.0883 | 3.2845 |
| **Pb206** | 0.0989 | 0      | 82 | -0.008 | 0.8031 | 2.097  | **Gd158** | 5.019  | 6.4615 | 64 | 0.271  | 0.0795 | 3.2883 |
| **Cd102** | 0.256  | 1.3333 | 48 | 0.053  | 0.7765 | 2.1088 | **Er166** | 5.822  | 7.4667 | 68 | 0.283  | 0.0806 | 3.2887 |
| **Kr88**  | 0.465  | 1.6    | 36 | 0.062  | 0.7753 | 2.1203 | **Th234** | 7.92   | 5.5385 | 90 | 0.215  | 0.0496 | 3.2896 |
| **Mo100** | 0.53   | 4      | 42 | 0.244  | 0.5356 | 2.1212 | **Sm156** | 7.2    | 6      | 62 | 0.279  | 0.0759 | 3.2903 |
| **Pd100** | 0.347  | 2      | 46 | 0.088  | 0.6656 | 2.1279 | **Hf178** | 4.736  | 6.6667 | 72 | 0.278  | 0.0932 | 3.2906 |
| **Pt176** | 2.55   | 3.2    | 78 | 0.171  | 0.264  | 2.1367 | **W182**  | 4.075  | 5.5385 | 74 | 0.259  | 0.1001 | 3.2908 |
| **Ru98**  | 0.401  | 2.4    | 44 | 0.115  | 0.6524 | 2.1424 | **U232**  | 9.91   | 5.8333 | 92 | 0.207  | 0.0476 | 3.2912 |
| **Sm148** | 0.712  | 3      | 62 | 0.161  | 0.5503 | 2.1448 | **Yb170** | 5.734  | 7.2    | 70 | 0.295  | 0.0843 | 3.2929 |
| **Se74**  | 0.357  | 3.75   | 34 | -0.25  | 0.6348 | 2.1476 | **Dy162** | 5.315  | 7.4667 | 66 | 0.281  | 0.0807 | 3.2936 |
| **Xe132** | 0.466  | 2      | 54 | 0      | 0.6677 | 2.1571 | **U234**  | 10.25  | 6.1538 | 92 | 0.215  | 0.0435 | 3.2956 |
| **Se70**  | 0.29   | 3.4286 | 34 | -0.307 | 0.9446 | 2.1575 | **Dy164** | 5.616  | 8      | 66 | 0.292  | 0.0734 | 3.3005 |
| **Zn76**  | 0.145  | 1.3333 | 30 | 0.142  | 0.5987 | 2.1655 | **Gd162** | 5.51   | 7.4667 | 64 | 0.291  | 0.0716 | 3.3017 |
| **Cr54**  | 0.0865 | 1.3333 | 24 | 0.18   | 0.8349 | 2.1847 | **Gd160** | 5.184  | 7      | 64 | 0.28   | 0.0753 | 3.3022 |
| **Gd152** | 1.655  | 4.2    | 64 | 0.207  | 0.3443 | 2.194  | **U236**  | 11.06  | 6.4286 | 92 | 0.215  | 0.0452 | 3.3039 |
| **Ti52**  | 0.0603 | 1      | 22 | 0      | 1.0497 | 2.2079 | **U238**  | 12.06  | 6.6667 | 92 | 0.215  | 0.0449 | 3.3039 |
| **Rn220** | 1.875  | 2.6667 | 86 | 0.111  | 0.241  | 2.2146 | **Yb172** | 6.024  | 7.5    | 70 | 0.296  | 0.0787 | 3.3053 |
| **Xe140** | 0.329  | 2      | 54 | 0.116  | 0.3767 | 2.2148 | **Hf180** | 4.647  | 6.4286 | 72 | 0.279  | 0.0933 | 3.3065 |
| **Zr86**  | 0.157  | 2.8571 | 40 | 0.053  | 0.7518 | 2.2169 | **Pu242** | 13.71  | 7.7647 | 94 | 0.224  | 0.0445 | 3.3071 |
| **Kr74**  | 0.614  | 4.4444 | 36 | 0.4    | 0.4556 | 2.2239 | **Pu240** | 13.03  | 7.5    | 94 | 0.223  | 0.0428 | 3.3087 |
| **Sr84**  | 0.292  | 2.8571 | 38 | 0.053  | 0.7932 | 2.2286 | **Cm248** | 14.61  | 9.1    | 96 | 0.235  | 0.0434 | 3.3088 |
| **Ge68**  | 0.1242 | 2.6667 | 32 | -0.275 | 1.016  | 2.2323 | **Er168** | 5.786  | 7.875  | 68 | 0.294  | 0.0798 | 3.3092 |
| **Dy154** | 2.437  | 4.3636 | 66 | 0.207  | 0.3343 | 2.2325 | **Yb174** | 5.94   | 7.7647 | 70 | 0.287  | 0.0765 | 3.31   |
| **Zn68**  | 0.1203 | 1.6667 | 30 | -0.156 | 1.0774 | 2.2438 | **Yb176** | 5.189  | 7.5    | 70 | 0.278  | 0.0821 | 3.31   |
| **Xe130** | 0.603  | 2.4    | 54 | -0.113 | 0.5361 | 2.2471 | **Er170** | 5.85   | 8.2353 | 68 | 0.296  | 0.0786 | 3.3102 |
| **Ti46**  | 0.0951 | 1.3333 | 22 | 0      | 0.8893 | 2.2601 | **Pu238** | 12.32  | 7.2    | 94 | 0.215  | 0.0441 | 3.3114 |
| **Ti44**  | 0.068  | 1      | 22 | 0      | 1.0831 | 2.2661 | **Cm244** | 14.79  | 8.5556 | 96 | 0.234  | 0.043  | 3.3131 |
| **Cd104** | 0.322  | 1.5    | 48 | 0.089  | 0.658  | 2.2676 | **Cm246** | 14.99  | 8.8421 | 96 | 0.234  | 0.0429 | 3.314  |
| **Ge66**  | 0.1393 | 2.4    | 32 | 0.229  | 0.957  | 2.2714 | **Cf252** | 16.7   | 10.182 | 98 | 0.236  | 0.0457 | 3.3189 |
| **Ru100** | 0.4927 | 3      | 44 | 0.161  | 0.5395 | 2.2734 | **Cf250** | 16     | 9.9048 | 98 | 0.245  | 0.0427 | 3.321  |

Table (1): The data set used to perform the PCA classified in ascending way according to their $R_{4/2}$ ratio.

## 3- Method of Analysis

Principal Component Analysis (PCA) is a powerful and widely used technique to analyze data of multivariate structure [14-17]. PCA analyzes a data table representing observations (nuclei) described by several dependent variables, which are, in general, inter-correlated. The goals of PCA are to [18] (a) extract the most important information from the data set, (b) compress the size of the data set by keeping only this important information, (c) simplify the description of the data set, and (d) analyze the structure of the observations and the variables. In order to achieve these goals, PCA computes new variables called *principal components* which are obtained as linear combinations of the original variables. The values of these new variables for the observations are called *factor scores,* and these factors scores can be interpreted geometrically as the projections of the observations onto the principal components [18].

The first step of PCA calculates the covariance matrix, but in current study the variables ($R_{4/2}$, $Z$, $\beta_2$, $P$, $2^+$, and $B(E2)$) having different variances, and units of measurements. Consequently, the correlation matrix is the standard method. If we have $I$ nuclei described by $J$ variables as in table (1), so we have $I \times J$ matrix, $\mathbf{X}$, whose elements $x_{i,j}$. We have to standardize the data set by subtracting off the mean, and dividing by the standard deviation of each column of $\mathbf{X}$, this gives us standardized data matrix say, $\dot{\mathbf{X}}$.

The correlation between variables, say $x_i$, and $x_j$ is calculated as follows,

$$corr(x_i, x_j) = \frac{1}{n-1} \sum_{i,j=1}^{n} \left( \frac{x_i - \bar{x}_i}{\sigma_{x_i}} \right) \left( \frac{x_j - \bar{x}_j}{\sigma_{x_j}} \right); \qquad (1)$$

where $\bar{x}_{i,j}$, and $\sigma_{x_{i,j}}$ are the mean value, and the standard deviation of variables $x_{i,j}$ respectively. Or simply in the language of matrix algebra, the correlation matrix $\mathbf{C}$ of matrix $\mathbf{X}$, that describes all relationships between pairs of measurements is defined as

$$\mathbf{C} = \frac{1}{n-1} \mathbf{D}^{-\frac{1}{2}} . X . X^T . \mathbf{D}^{-\frac{1}{2}}, \qquad (2)$$

where $\mathbf{D}^{-\frac{1}{2}} = [1/\sigma_{x_j}]$ is a diagonal matrix. With the correlation matrix, the eigenvectors and eigenvalues are calculated, and then the eigenvalues are sorted in descending order. This gives us the components in order of significance. The eigenvector with the highest eigenvalue is the most dominant principle component of the data set (PC1). It, (PC1), expresses the most significant relationship between the data dimensions.

The PC1 is required to have the largest possible variance. The second component (PC2) is computed under the constraint of being orthogonal to the first component and to have the second largest variance. The observations (nuclei) can be represented in the PC space by their factor scores. This can be done by singular value decomposition (SVD) of standardized data matrix $\dot{\mathbf{X}}$ [19]:

$$\dot{\mathbf{X}} = \mathbf{P} \Delta \mathbf{Q}^T \qquad (3)$$

where $\mathbf{P}$ is the matrix of left singular vectors, $\mathbf{Q}$ is the matrix of right singular vectors, and $\Delta$ is the diagonal matrix of singular values (eigenvalues). The matrix of factor scores, $\mathbf{F}$, is defined as:

$$\mathbf{F} = \mathbf{P}\Delta = \Delta \mathbf{Q}^T \mathbf{Q} = \dot{\mathbf{X}} \mathbf{Q} . \qquad (4)$$

The matrix $\mathbf{Q}$ gives the coefficients of the linear combinations used to compute the factors scores. This matrix can also be interpreted as a projection matrix because it gives the values of the *projections* of the observations on the principal components.

The variables can be plotted as points in the component space using their inter-correlation as coordinates. This correlation is called a *loading* in PCA terms [18] (the blue color in figure (1)). The sum of the squared loadings for a variable is equal to one; consequently, the loadings will be positioned on a circle which is called the *circle of correlations*. The closer variable is to the circle of correlations, the better we can reconstruct this variable from the first two components (and the more important is to interpret these components); the closer to the center of the plot a variable is, the less important is for the first two components [18].

A useful tool to represent and interpret the results of PCA is called biplot, which represents the correlations between nuclei and their variables in the space of the first two principal components. Practically, the PCA is performed by standard packages of statistical software without dealing of complicated steps to build the biplot.

## 4- Results and Discussions

Recalling that every measured physical quantity is combined by a degree of uncertainty, we should take into consideration the error analysis. The PCA is used in two main ways:

(1) The reduction, where focus is given on the extracted components to simplify and/or model data of high dimensions. The error analysis here is highly required. A recent study of Dobaczewski *et al.* [20] provides suggestions on uncertainty quantification of nuclear structure models.

(2) The visualization, where attention is given on the way of variables and observations are inter-correlated in the biplot. In present study, we focus on the visualization of data set, search for clusters, and the way of different nuclei which respond to same nuclear variables.

Most error treatments are used in data reduction to estimate uncertainties of extracted parameters from PCA. Recently, the authors of [21] studied the impact of uncorrelated measurement error on extracted eigenvalues and eigenvctors of PCA. They suggest that the impact is negligible when these component scores are used for visualizing data.

The uncertainties of variables considered in current study are different in weight from one variable to another and even from each nucleus to the other (heterogeneous); in addition, they are correlated. Apparently, analysis of this kind of error has not been studied so far. It is, therefore, hoped that findings of [21] should also apply to our study.

First, we start by applying two tests [22] on the correlation matrices of each $R_{4/2}$ ratio:

1- *Kaiser-Meyer-Olkin (KMO) Measure of Sampling Adequacy*: This measure varies between 0 and 1, and values closer to 1 are better. A value below 0.5 is not acceptable.

2- *Bartlett's Test of Sphericity* : This tests the null hypothesis that the correlation matrix is an identity matrix (variables are not correlated).

Taken together, they provide a minimum standard which should be passed before PCA is conducted. These tests are found in commercial software packages, our correlation matrices passed them.

Figure (1) shows the extracted first two principal components of the correlation matrix of data set of table (1). A detailed analysis of each $R_{4/2}$ range is given in figure (2- a,b,c). Detailed calculations of the 1.95≤ $R_{4/2}$ < 2.05 range is given, since it contains only 10 nuclei to simplify the presentation. Other ranges are calculated in the same way. The standardized data matrix $\dot{X}$ is given in the first six columns of table (2) by subtracting the mean then dividing by the standard deviation of each column.

The correlation matrix is given in the first six columns of table (3). By diagonalizing the correlation matrix, the eigenvalues $\lambda_i$ are calculated in the seventh column. The first two eigenvalues represent nearly 82% of total $\lambda_i$ (($\lambda_1+\lambda_2$)/$\sum \lambda_i$), *i.e.*, the extracted first two principal components PC1, and PC2 in figure (2-a) covers about 82% of the data set. The last two columns of table (3) represent the loadings of matrix Q corresponding to PC1, and PC2 of each variable represented by blue color in figure (2-a) scaled on the top and right axes. For example, the position of the line B(E2) is given by the loading numbers 0.51 and -0.2 on the far right of the first line of table (3). By multiplying these loadings by the standardized matrix $\dot{X}$, we get the factor scores of eq. (4) in the last two columns in table (2) of each nucleus, represented in figure (2-a) on bottom and left axes. For illustration, the point for $^{62}$Ni nucleus is drawn in Fig (2-a) at the coordinates (scores) -2.76 and -2.19 given in the columns called scores on the third line of table 2.

The following points can be extracted from figures (1), and (2-a,b,c):

1- The variables under investigation may be grouped into two categories according to their effect on collective behavior of nuclei. Generally speaking, collective behavior of nuclei of collective parameters such as *B(E2)*, *P*, *β₂*, *R₄/₂,* and Z, where in general increases as the parameter values increase – in case of Z parameter, it only applies from a closed shell to around mid-shell-. On the contrary, the 2⁺ energy states increases by decreasing collectivity. Subsequently, we expect negative correlation between 2⁺ energy states loadings and other collective parameters. This finding is indicated clearly in figure (1).

2- The 2⁺ loadings have a very small projection on the PC1 axis and this is confirmed in figures 2-a,b,c. Therefore, PC1 values can be used as a sign of collectivity, where the highest value of PC1 corresponds to $^{252}$Cf of highest collectivity. Whereas, the lowest value (negative) of PC1 corresponds to $^{62}$Ni of least collectivity since it has spherical structure due to the closed shell Z=28.

3- The proposed nuclei of *SU*(3) dynamical symmetry are all found on the right hand side of figure (1). The SU(3) nuclei have the highest values of collective parameters ( see table (1) for details) while they have the least values (about an order of magnitude less) of 2⁺ energy states compared to the U(5), and the range 2.05≤ $R_{4/2}$ < 2.30 nuclei. Consequently, the SU(3) nuclei are positively correlated (closer) to the collective parameters, and negatively correlated (on the other side) to 2⁺ energy states loadings.

4- In figure (1), the SU(3) nuclei are grouped into two clusters. The actinides of Z> 82 isotopes are lying in a line between the B(E2) and *Z* loadings in the right lower quadrant. In addition to a tight cluster in the upper right quadrant of the rest of *SU*(3) nuclei, including lanthanides in addition to Hf, and W isotopes which are correlated to *β₂* loadings. (N.B. for simplicity, we will call this cluster lanthanides other not stated), 5-  The enhanced correlation between actinides cluster and B(E2), and Z loadings comes from the way these parameters change with nuclei under investigation. From table (1), we can observe that, the B(E2) values span the range from 3.7 to 7.2 for lanthanides, while the same parameter (B(E2)) takes the range from 7.9 to 16.7 for actinides. The higher values of B(E2) of actinides will enhance the positive correlation (closer) compared to lanthanides. On the other hand, as the range of change of parameter space increases, the distribution of nuclei positions (scores) will increase. For instance, the range of change of B(E2) is $\Delta B(E2)=16.7-7.9=8.8$ for actinides, while it is $\Delta B(E2)=7.2-3.7=3.5$ for lanthanides. As noticed, the scores of actinides will span a wider area (a line in figure (1)) than lanthanides (tight cluster). The same is true of Z parameter, where Z spans from 90 to 98 (higher values →higher correlation), and from 60 to 74 for actinides and lanthanides respectively.

6-  The effect of $β_2$ parameter (loadings) contradicts that of B(E2), and Z loadings as seen in figure (1). The $β_2$ spans the range of 0.19 to 0.24 and from 0.24 to 0.29 for actinides, and lanthanides. As a result, lanthanides will be more highly correlated to $β_2$ parameter than that of actinides. The range of change of $β_2$ is $\Delta β_2 ≈0.05$ is similar for both clusters, consequently, we predict no effect on the shape of each cluster compared to $\Delta B(E2)$.

7-  The $P$-factor and $R_{4/2}$ ratio loadings in figure (1) are lying in-between the two clusters of SU(3) nuclei without significant correlation to one of them and this is confirmed in figure (2-c); this trend may be interpreted as follows. From table (1), a unit change in the $P$-factor, say from 5 to 6; there are 10 nuclei, 5 of actinides and 5 of lanthanides and W isotopes. The same is true for $R_{4/2}$ ratio, if we take a change from, say 3.30 to 3.31, we get 11 nuclei, 5 of actinides, and the rest 6 of lanthanides and Hf isotope. The conclusion here is that, the change of $P$-factor and $R_{4/2}$ ratio will not differentiate between the nuclei of each cluster. Accordingly, the correlation will be similar for the two clusters.

8-  Combining the previous effect of $2^+$, B(E2), Z, $β_2$, P, and $R_{4/2}$ together on the SU(3) nuclei in the presence of the rest of data set gives us the pattern of SU(3) nuclei in figure (1).

9-  For SU(3) nuclei in figure (2-c), the $2^+$ energy states are spanning the range of 0.04 to 0.05 MeV and 0.07 to 0.11 MeV for actinides and lanthanides, respectively. This will enhance the positive correlation between lanthanides and $2^+$ loadings on the left side of figure (2-c). This effect was not clear in figure (1) because the presence of U(5), and the range $2.05≤ R_{4/2} < 2.30$ nuclei as discussed in point (3).

10- The position of each nucleus (scores) and parameters (loadings) in figures (1), and (2-a,b,c) are extracted from the correlation matrix of each range of nuclei under investigation. Consequently, any change of the number of nuclei or parameters participating in the calculation of the correlation matrix will affect extracted values of the corresponding scores and loadings.

11- The previous points can help us to understand the difference in pattern of SU(3) candidate nuclei between figure (1) and figure (2-c). The actinides having enhanced positively correlation to B(E2), and Z more than $β_2$, and $2^+$ parameters, while the opposite is true for lanthanides. On the other hand, both clusters have similar correlation to $P$, and $R_{4/2}$ parameters. Combining these effects together may explain why actinides and lanthanides are distributed in nearly two parallel diagonal lines in figure (2-c).

12- The position of each nucleus in the biplot of figures (1), and (2-a,b,c) is given as an average correlation (close or far) of all parameters under study. For example, in figure (2-c), $^{230}$Th -found in the right lower quadrant- has the highest $2^+$ energy state compared to other actinides, subsequently, it's the closest to the left side where the $2^+$ loading exist. The same nucleus has the lowest $\beta_2$, $R_{4/2}$, P, and Z values, while the second lowest value of B(E2). Consequently, $^{230}$Th takes the farthest position from these parameters.

13- The scattered distribution of $U(5)$, and the range $2.05 \leq R_{4/2} < 2.30$ nuclei cannot make us extract a clear conclusion from figure (1).

14- Figure 2-a shows that the $\beta_2$ loadings are the affecting parameter on the $U(5)$ nuclei. Although, The $R_{4/2}$ loading is in the same direction, it is less correlated to the principal components (PC1, and PC2) (shorter).

15- Figure 2-b clarifies that for the $R_{4/2}$, and $\beta_2$ loadings highly affect the $2.05 \leq R_{4/2} < 2.30$ range, whether by positive or negative correlation, while the P-factor is the weaker.

|  | *B(E2)* | *P* | *Z* | *β₂* | *R₄/₂* | *2⁺* | scores | |
|---|---|---|---|---|---|---|---|---|
| **Sr⁹⁰** | -0.83 | 0.203 | -0.69 | 1.003 | -1.06 | 0.293 | *-0.97* | *0.3* |
| **Te¹¹⁸** | 0.302 | 0.385 | 0.042 | -1.45 | -1.02 | -0.56 | *0.88* | *-1.41* |
| **Ni⁶²** | -0.86 | -2.53 | -1.2 | -0.83 | -1.02 | 1.578 | *-2.76* | *-2.19* |
| **Hg¹⁸⁶** | 2.233 | 0.452 | 1.496 | -1.24 | -0.94 | -1.31 | *3.02* | *-1.48* |
| **Rn²¹⁸** | 1.074 | 1.407 | 1.807 | 0.843 | 0.078 | -1.62 | *2.78* | *1.17* |
| **Te¹²⁸** | -0.16 | -0.07 | 0.042 | 0.352 | 0.092 | -0.04 | *-0.12* | *0.29* |
| **Zn⁷⁰** | -0.71 | 0.203 | -1.1 | 0.904 | 0.342 | 0.492 | *-1.26* | *0.95* |
| **Zr⁸⁸** | -0.87 | 0.203 | -0.58 | 1.003 | 0.527 | 1.142 | *-1.44* | *1.16* |
| **Te¹²⁶** | 0.07 | 0.094 | 0.042 | -0.94 | 1.483 | -0.33 | *0.35* | *0.24* |
| **Xe¹³⁴** | -0.26 | -0.34 | 0.145 | 0.352 | 1.517 | 0.351 | *-0.49* | *0.97* |

Table (2): The standardized data matrix $\dot{\mathbf{X}}$ of $U(5)$ nuclei is given in the first six columns by subtracting the mean and dividing by the standard deviation of each column. The last two columns represent the factor scores of eq. (4) for each nucleus.

|  | *B(E2)* | *P* | *Z* | *β₂* | *R₄/₂* | *2⁺* | λ | loadings | |
|---|---|---|---|---|---|---|---|---|---|
| **B(E2)** | 1 | 0.493 | 0.904 | -0.46 | -0.18 | -0.88 | **3.362** | *0.51* | *-0.2* |
| **P** | 0.493 | 1 | 0.639 | 0.279 | 0.159 | -0.76 | **1.53** | *0.4* | *0.45* |
| **Z** | 0.904 | 0.639 | 1 | -0.18 | 0.027 | -0.91 | **0.772** | *0.52* | *0.05* |
| **β₂** | -0.46 | 0.279 | -0.18 | 1 | 0.28 | 0.231 | **0.247** | *-0.15* | *0.67* |
| **R₄/₂** | -0.18 | 0.159 | 0.027 | 0.28 | 1 | 0.064 | **0.056** | *-0.04* | *0.55* |
| **2⁺** | -0.88 | -0.76 | -0.91 | 0.231 | 0.064 | 1 | **0.033** | *-0.53* | *-0.04* |

Table (3): The correlation matrix is given in the first six columns of $U(5)$ nuclei. The eigenvalues $\lambda_i$ of the correlation matrix is given in the seventh column. The last two columns represent the loadings of matrix Q corresponding to first two eigenvalues.

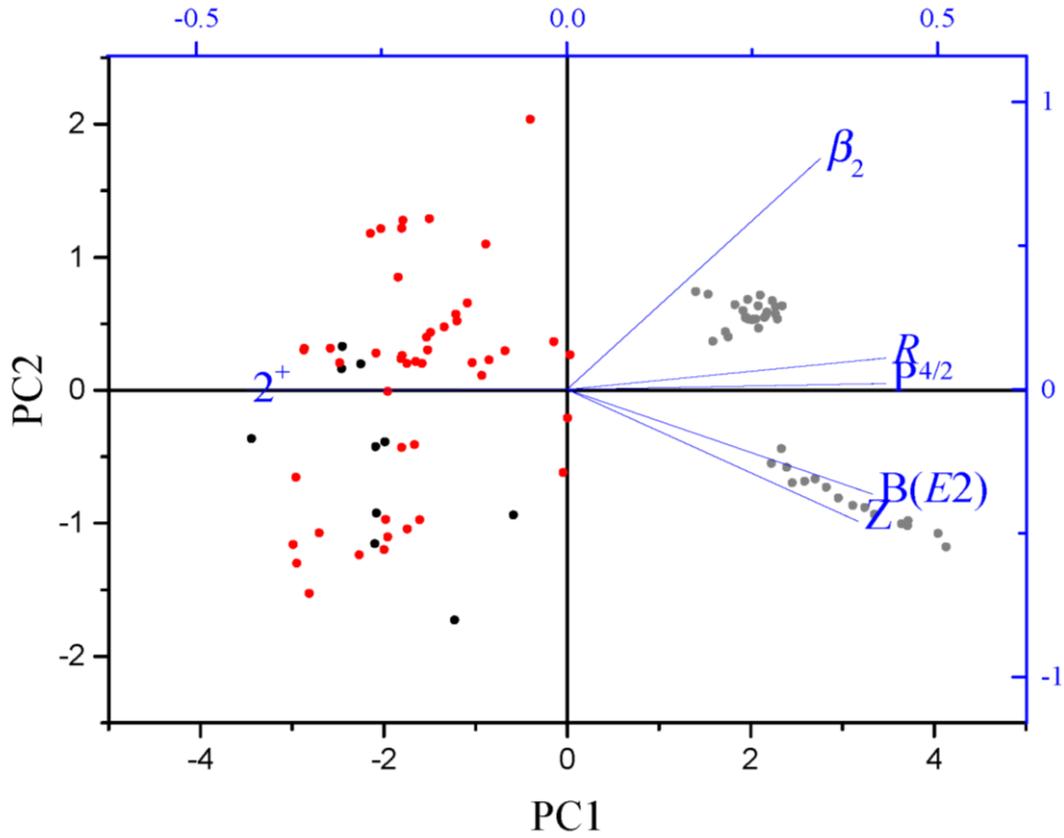

Figure (1) (color on line): shows the first two principal components of all the data set of table (1). Blue color represents the loadings of the six variables scaled on the top and right axes. Black, red, and gray dots correspond to the $U(5)$, $2.05 \leq R_{4/2} < 2.30$ range, and $SU(3)$ nuclei scores, respectively scaled on the bottom and left axes.

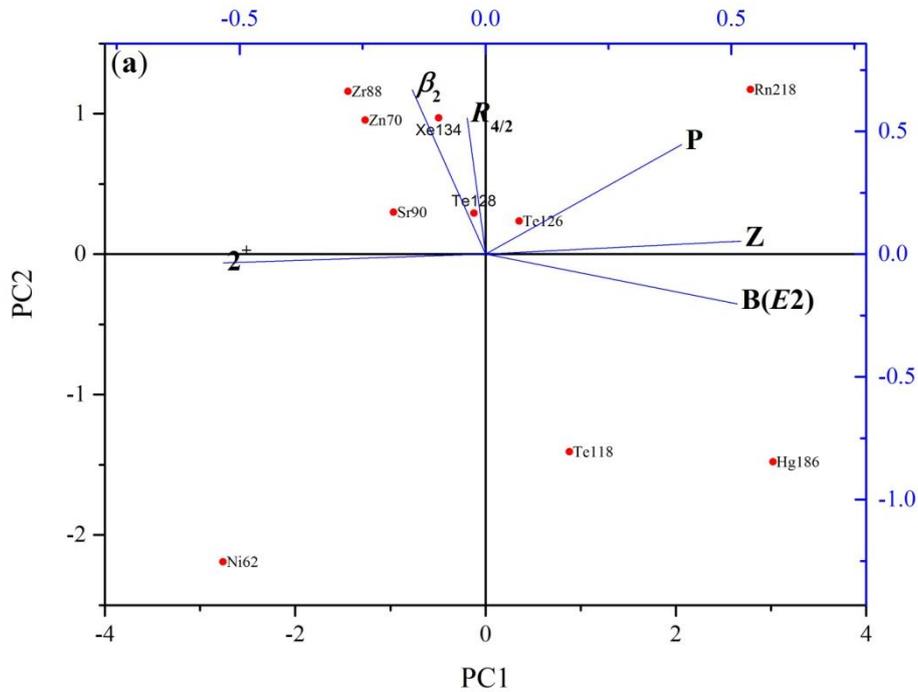

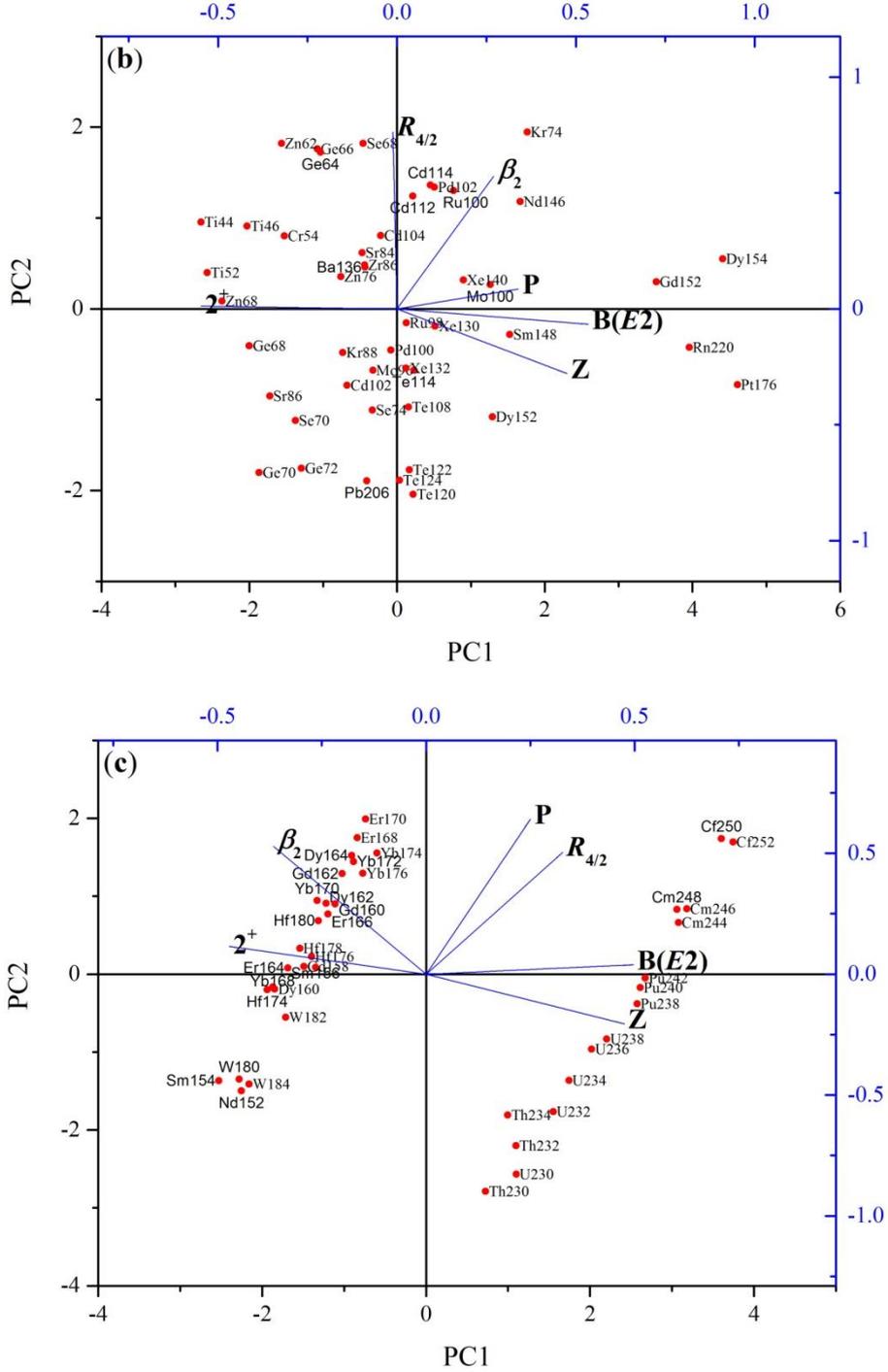

Figure (2-a,b,c): the same as figure (1), but for $U(5)$, $2.05 \leq R_{4/2} < 2.30$ range, and $SU(3)$ nuclei, respectively.

## 5- Conclusion

The understanding of collective behavior of atomic nuclei is still an important field of research. Through this paper we have tried to answer the question of simultaneous effect of different nuclear parameters on nuclear structure. The principal component analysis is used throughout this study to visualize this effect. The results suggest that in the presence of other parameters, the $R_{4/2}$ ratio loses its advantages to characterize the IBM dynamical symmetries candidate nuclei. We have also found that the PC1 of the correlation matrix of the data set classifies the *SU*(3) nuclei into two clusters of actinides, and lanthanides. This finding can be used for a wider range of data to identify them based on their position in the biplot. Finally, we can conclude that the use of principal component analysis within nuclear structure can give a simple method to visualize the effect of different parameters on nuclear structure.